\documentclass[runningheads]{llncs}
\usepackage[T1]{fontenc}
\usepackage[utf8]{inputenc} 
\usepackage{orcidlink}
\usepackage{graphicx}
\usepackage{amsmath}
\usepackage{amssymb}
\usepackage{subcaption}
\usepackage{mathptmx}

\graphicspath{ {figures/} }
\usepackage{color}

\usepackage{url}
\usepackage{braket}
\usepackage{mathtools, amsfonts, bm, dsfont}
\usepackage{flushend}
\usepackage[capitalize]{cleveref}
\usepackage[numbers]{natbib}
\usepackage[dvipsnames]{xcolor}

\begin{document}
\title{Circuit-Level Analysis of Quantum Error Correction in Second-Generation Quantum Repeaters}
\titlerunning{Quantum Error Correction in Second-Generation Quantum Repeaters}
%
\author{Suraj Vishwanath\inst{1}\orcidlink{0009-0006-4750-200X} \and
Jonas Hawellek\inst{2} \and
Roberto Ferrara\inst{2}\orcidlink{0000-0002-1991-3286} \and
Christian Deppe\inst{2,3}\orcidlink{0000-0002-2265-4887}}
\authorrunning{S. Vishwanath, J. Hawellek, R. Ferrara, C. Deppe}
%
\institute{Department of Physical Sciences, Indian Institute of Science Education and Research (IISER), Mohali, Punjab 140306, India \and
Technische Universität Braunschweig, Institute for Communications Technology, Braunschweig, Germany \and 6G-life, 6G Research Hub, Germany\\
\email{surajv@vt.edu, jonas.hawellek@tu-braunschweig.de, roberto.ferrara@tu-braunschweig.de, christian.deppe@tu-braunschweig.de}}
\maketitle              

\begin{abstract}
We present a circuit-level simulation of second-generation quantum repeaters with stabilizer-based quantum error correction.
We model a two-segment repeater link including encoding, syndrome extraction, decoding, and entanglement swapping, and compare the performance of the single-error-correcting Steane and five-qubit stabilizer codes under different noise scenarios.
To that extent, we consider noise sources with incoherent (depolarizing and dephasing) and coherent (systematic over-rotation) errors and incorporate them directly at the circuit level.
While the Steane code outperforms the five-qubit code at coherent errors, the five-qubit code achieves higher end-to-end fidelities under incoherent noise in the low-noise regime when a large number of physical qubits is affected by it.
At higher noise regimes, employing no code turns out to be the best option.
In mixed noise scenarios, we observe moderate cross-effects between the noise classes regarding the error correction performance.
We conclude that code selection in second-generation quantum repeaters must be adapted to the dominant physical noise mechanism and that circuit complexity and hardware-aware implementation options influence performance in coherent-error-dominated regimes, providing design insights for error correction in quantum repeaters.
\end{abstract}

\section{Introduction}

Quantum repeaters \cite{PhysRevLett.81.5932,  Jiang_Taylor_Nemoto_Munro_VanMeter_Lukin_2009, Loock2019ExtendingQL, Muralidharan2015OptimalAF, RevModPhys.83.33, Zwerger_Briegel_Dür_2014} are a key enabling technology for long-distance quantum communication, overcoming the exponential loss and noise limitations of direct transmission by distributing entanglement over multiple segments and connecting them via intermediate nodes.
First-generation quantum repeaters \cite{PhysRevLett.81.5932, Muralidharan2015OptimalAF, RevModPhys.83.33} combine heralded entanglement generation with entanglement purification to mitigate both loss errors and noise.
In order to reduce the amount of long-range classical communication required for purification, second-generation quantum repeaters \cite{Jiang_Taylor_Nemoto_Munro_VanMeter_Lukin_2009, Zwerger_Briegel_Dür_2014} employ quantum error correction (QEC) to mitigate noise.
While the theoretical principles of second-generation repeaters are well understood, several practical questions remain insufficiently explored. 
In particular, the interplay between different physical noise mechanisms and circuit-level implementations of QEC has not been systematically analyzed for minimal stabilizer codes.
Small-distance codes such as the $[[5,1,3]]$ five-qubit code and the $[[7,1,3]]$ Steane code both correct arbitrary single-qubit errors, yet they differ significantly in structure and circuit complexity. 
These structural differences can lead to markedly different behavior under different noise sources, especially when realistic gate decompositions and hardware constraints are taken into account.

In this work, we perform a circuit-level simulation of a second-generation quantum repeater link that enables a systematic comparison of such codes under realistic noise models. 
We distinguish between incoherent noise, modeled as depolarizing and phase-flip channels, and coherent noise, modeled as systematic over-rotation errors, for example due to clock drift.
To obtain representative results, we perform full density-matrix simulations incorporating both kinds of noise at the level of physical gate operations and transmission channels.
We model a two-segment repeater link including encoding, noisy transmission, syndrome extraction, decoding, and entanglement swapping, allowing end-to-end Bell-state fidelity to be evaluated under controlled noise regimes.

Our results demonstrate that code selection in second-generation repeaters is noise-regime dependent and that circuit implementation becomes a decisive factor in coherent-error-dominated settings. 

\section{Preliminaries}

\subsection{Quantum Repeaters}

\begin{figure}[t]
    \centering
    \includegraphics[width=0.75\linewidth]{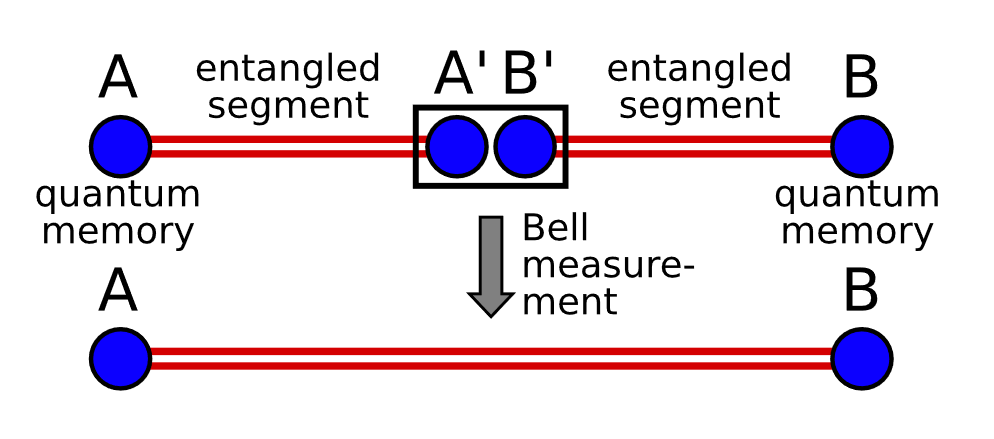}
    \caption{Schematic of a two-segment repeater link \cite{Loock2019ExtendingQL}. Neighboring nodes establish entanglement which is then connected via swapping operations.}
    \label{fig:two_segment_repeater}
\end{figure}

The central challenge of quantum communication is the distribution of high-fidelity entanglement over long distances in the presence of loss and operational noise. 
Quantum repeaters mitigate exponential transmission loss by dividing the total distance between distant nodes $A$ and $B$ into shorter segments, generating entanglement locally between neighboring nodes, and extending it via entanglement swapping \cite{PhysRevLett.81.5932,Loock2019ExtendingQL,Muralidharan2015OptimalAF}.
A schematic of a simple equal-length, two-segment repeater link is given in Figure~\ref{fig:two_segment_repeater}.
By introducing an intermediate repeater node, the link is divided into two equal segments for which we assume the same error models.

\begin{figure}[ht]
    \centering
    \includegraphics[width=\linewidth]{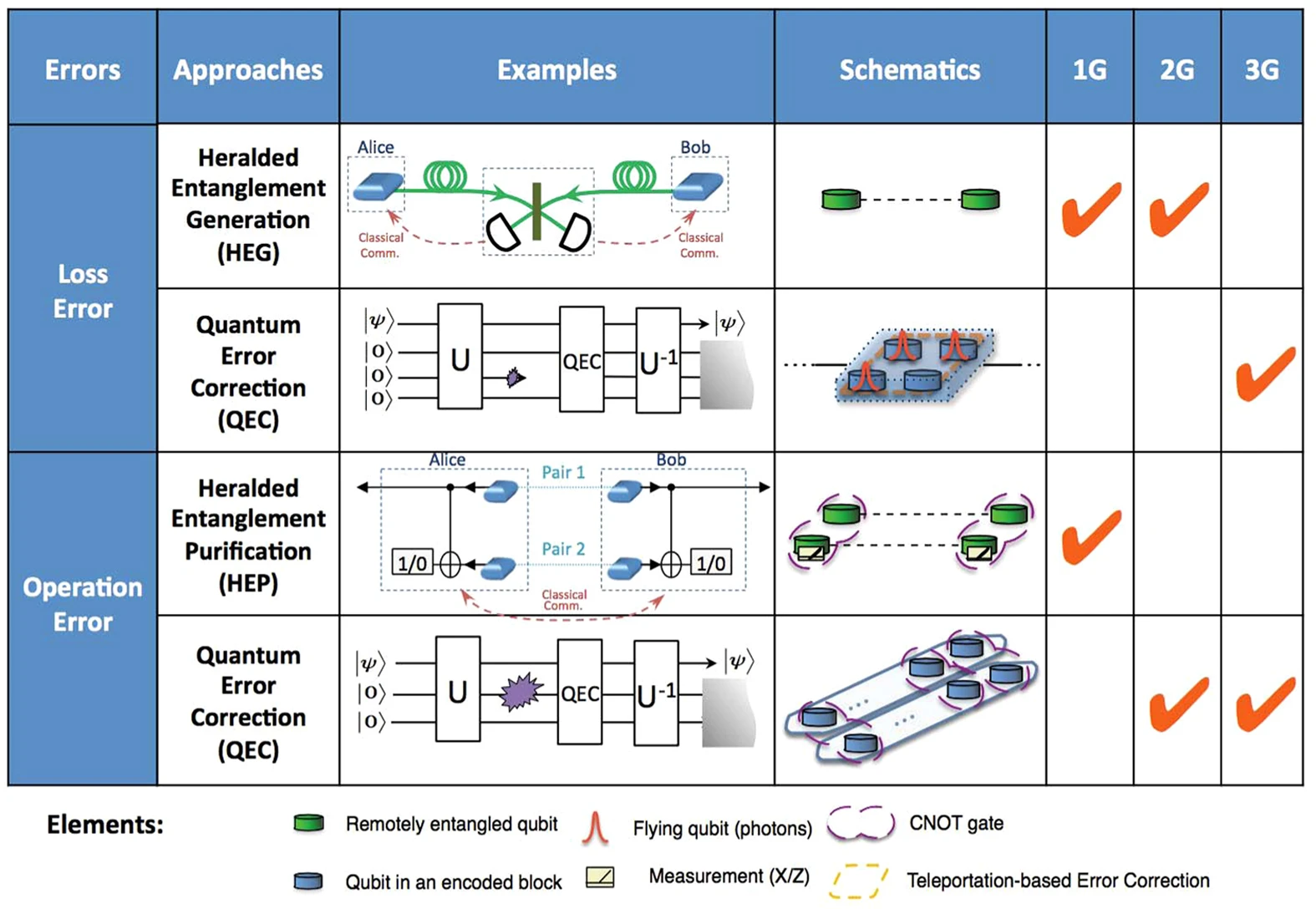}
    \caption{Conceptual comparison of repeater generations \cite{Muralidharan2015OptimalAF}. First generation relies on two-way entanglement distillation; second generation uses QEC for operational errors while maintaining heralding for loss; third generation aims for fully one-way, QEC-based operation. The schematics illustrate the physical resources involved, including entangled qubits, flying qubits, encoded qubit blocks, measurements, and CNOT-based purification.}
    \label{fig:repeater_generations}
\end{figure}

Following \cite{Muralidharan2015OptimalAF}, quantum repeaters are commonly categorized into three generations.
The three generations of quantum repeaters are classified according to whether they use heralded processes and/or quantum error correction for establishing elementary links and correcting errors.
In the following, we briefly discuss the differences between the generations, as illustrated in Fig~\ref{fig:repeater_generations}.

\paragraph{First generation (1G).}
First-generation repeaters \cite{PhysRevLett.81.5932, Muralidharan2015OptimalAF, RevModPhys.83.33} employ heralded entanglement generation (HEG) protocols to handle loss errors, where successful generation of entanglement between distant nodes is confirmed via classical communication.
To suppress operational errors, this generation utilizes heralded entanglement purification (HEP) \cite{RevModPhys.95.045006,PhysRevA.80.042308}, using local operations and classical communication (LOCC) to distill high-fidelity entangled states.
The two-way classical communication required incurs substantial latency at long distances.

\paragraph{Second generation (2G).}
Second-generation repeaters \cite{Jiang_Taylor_Nemoto_Munro_VanMeter_Lukin_2009, Zwerger_Briegel_Dür_2014} retain heralding to handle loss, but replace heralded purification by quantum error correction (QEC) to suppress operational errors \cite{PhysRevA.54.1098,Muralidharan2015OptimalAF,Chuang2020QuantumCA}.
Encoding Bell pairs into logical qubits and performing error correction reduces the observed noise on the shared entanglement without long-distance two-way signaling.
However, due to the HEG between adjacent nodes, classical communication is still necessary on each segment.

\paragraph{Third generation (3G).}
Third-generation repeaters aim to correct both loss and operational errors deterministically using QEC, further reducing the need for classical communication.
However, the required gate fidelities, coupling efficiencies, and resource overheads remain beyond current experimental capabilities in most platforms \cite{Muralidharan2015OptimalAF}.

In this paper, we focus on second-generation repeaters, which represent a realistic intermediate regime: heralding remains feasible for lossy links, while QEC can improve robustness against operational noise at moderate resource overhead.

\subsection{Quantum Error Correction}\label{sec:stabilizercodes}

Quantum error correction (QEC) protects quantum information from noise without violating the no-cloning principle.
Stabilizer codes provide a convenient algebraic framework for QEC \cite{PhysRevA.54.1098, gottesman1997stabilizer, PhysRevLett.77.793}, where $k$ logical qubits are encoded into $n$ physical qubits.
An $[[n,k,d]]$ stabilizer code is specified by a stabilizer group, a set of mutually commuting Pauli operators.
This set is an Abelian subgroup $S=\langle g_1,\ldots,g_{n-k}\rangle$ of the $n$-qubit Pauli group, where $-\mathds{1}\notin S$.
The code space is the common $+1$ eigenspace of all stabilizer generators:
\begin{equation}
    \mathcal{C}
    =
    \left\{
        \ket{\psi}:
        g_i\ket{\psi}=\ket{\psi}
        \ \text{for all } i
    \right\}.
\end{equation}
The distance $d$ is the minimum weight of a Pauli operator that acts nontrivially on a logical state without taking it out of the code space.
Consequently, a code of distance $d$ can detect up to $d-1$ errors and correct $ t\leq \frac{d-1}{2}$ arbitrary single-qubit errors.

Errors are identified through syndrome measurements of the stabilizer generators.
For a Pauli error $E$, the syndrome $\mathbf{s} = [s_1,\cdots,s_{n-k}]$ records whether $E$ commutes or anticommutes with each generator:
\begin{equation}
    g_iE = (-1)^{s_i}Eg_i, \qquad s_i\in\{0,1\}.
\end{equation}
The resulting syndrome $\mathbf{s}$ identifies the corresponding error class, without revealing the encoded quantum information.
In an implementation, ancilla qubits are coupled to the data qubits to measure the stabilizers, with the ancilla measurement outcomes storing the associated syndrome.
A recovery operation is then selected according to the measured syndrome.

In the following, we focus on two distance-three stabilizer codes that encode a single logical qubit and correct arbitrary single-qubit errors: the $[[5,1,3]]$ five-qubit code and the $[[7,1,3]]$ Steane code.
The five-qubit code is the smallest possible single-error-correcting quantum code.

\begin{itemize}
    \item \textbf{Five-qubit code} $[[5,1,3]]$ \cite{Laflamme1996PerfectQE}, with stabilizer generators
    \[
        \begin{aligned}
            S_1 &= XZZXI,\quad
            S_2 &= IXZZX,\quad
            S_3 &= XIXZZ,\quad
            S_4 &= ZXIXZ.
        \end{aligned}
    \]
    \item \textbf{Steane code} $[[7,1,3]]$ \cite{PhysRevLett.77.793}, with stabilizer generators
    \[
        \begin{aligned}
            S_1 &= I I I X X X X,\quad
            S_2 &= I X X I I X X,\quad
            S_3 &= X I X I X I X,\\
            S_4 &= I I I Z Z Z Z,\quad
            S_5 &= I Z Z I I Z Z,\quad
            S_6 &= Z I Z I Z I Z.
        \end{aligned}
    \]
\end{itemize}

For the five-qubit code, the syndrome consists of four measurement outcomes, whereas the Steane code produces six outcomes.
These binary results are mapped to the corresponding single-qubit errors.
For example, a $Z$ error on the first qubit is represented by $E_5=ZIIII$ for the five-qubit code and by $ E_7=ZIIIIII $ for the Steane code.
Using the generator ordering given above, these errors produce the syndromes $\boldsymbol{s}_5=1010$, and $\boldsymbol{s}_7=001000$, respectively.
In the five-qubit code, $E_5$ anticommutes with $S_1$ and $S_3$, while in the Steane code, $E_7$ anticommutes only with $S_3$.
The corresponding stabilizer measurements therefore change sign for precisely these generators, allowing the decoder to identify the error and apply the corrective operation $Z$ to the first physical qubit.

\subsection{Quantum Circuit Transpilation}\label{sec:transpilation}

Quantum repeater protocols are naturally specified using abstract, hardware-agnostic gate sets. In contrast, physical quantum devices expose restricted, calibrated native gate sets and limited qubit connectivity \cite{Fan2025CalibratingQG}. Transpilation bridges this gap by converting an abstract circuit into an equivalent hardware-executable circuit while (i) respecting device constraints and (ii) optimizing the implementation to reduce decoherence and accumulated gate errors.

Generally, transpilation does not change the logical circuit and thus does not interfere with the noise model.
In this work, however, we also consider a model of coherent noise on the gates themselves.
In that case, different implementations cannot be expected to be equivalent anymore.

A convenient intermediate form for compiling single-qubit gates into device-native operations is the decomposition of them into rotation gates.
Single-qubit rotation gates are defined by
\begin{equation}\label{eq:defRotation}
    R_P(\phi) = e^{-i \phi P/2} \;,
\end{equation}
where $P$ is a unitary Hermitian operator defining the rotation axis.
For $P\in\{X,Y,Z\}$, we obtain rotations around the standard axes, and with $\phi=\pi$ we obtain the Pauli operators.

\paragraph{Single-qubit decomposition.}
Any single-qubit unitary $U$ can be decomposed (up to global phase) into $Z$--$Y$--$Z$ rotations \cite{Chuang2020QuantumCA}:
\begin{equation}
    U = U_3(\theta,\phi,\lambda) = e^{i \frac{\phi+\lambda}{2}} R_{z}(\phi)\, R_{y}(\theta)\, R_{z}(\lambda).
    \label{eq:zyz_decomposition}
\end{equation}
This widely used parameterized gate $U_3(\theta,\phi,\lambda)$ has matrix representation
\begin{equation}
    U_3(\theta, \phi, \lambda)=
    \begin{bmatrix}
        \cos(\theta/2) & -e^{i\lambda}\sin(\theta/2)\\
        e^{i\phi}\sin(\theta/2) & e^{i(\phi+\lambda)}\cos(\theta/2)
    \end{bmatrix},
    \label{eq:u3_matrix}
\end{equation}
and provides a gate definition on basic operations as an intermediate step towards hardware-aware implementation.
For controlled gates, we extend this definition to
\begin{equation}
    CU_3(\theta, \phi, \lambda)= \ket{0}\bra{0} \otimes I + \ket{1}\bra{1} \otimes U_3(\theta, \phi, \lambda) \;, 
    \label{eq:cu3_matrix}
\end{equation}
decomposing the gate into a combination of controlled rotation gates.

\subsection{Noise Models}

In the following, we consider two broad noise classes for operational errors at the circuit level introduced through imperfect gates, decoherence, and environmental interactions: incoherent and coherent noise.

\subsubsection{Incoherent noise.}\label{sec:incoherentnoise}
Incoherent noise captures stochastic interactions with the environment that map pure states to mixed states. 
We consider standard single-qubit channels:
\begin{itemize}
    \item \textbf{Depolarizing noise}:
    \begin{equation}
        \mathcal{E}_{\mathrm{dep}}(\rho)= (1-p)\,\rho + \frac{p}{3}\bigl(X\rho X + Y\rho Y + Z\rho Z\bigr),
        \label{eq:depolarizing_channel_paper}
    \end{equation}
    with $p\in[0,1]$ being the depolarizing probability.
    The depolarizing noise, commonly defined as in \cref{eq:depolarizing_channel_paper}, is maximal for $p=\frac{3}{4}$ as this results in a maximally mixed state.
    This noise models the complete loss of state information and is therefore considered a worst-case assumption for transmission channels if the environmental interactions are unknown.
    \item \textbf{Phase-flip noise}:
    \begin{equation}
        \mathcal{E}_{\mathrm{ph}}(\rho)= (1-p)\,\rho + p\,Z\rho Z,
        \label{eq:dephasing_channel_paper}
    \end{equation}
    with $p\in[0,1]$ denoting the phase-flip probability.
    While this is the definition of phase-flip noise, another common and closely related term is dephasing noise.
    Dephasing does not include phase flips with $p>0.5$ such that the maximum dephasing noise is a randomized phase-flip with $p=0.5$, resulting in the maximum degradation of the quantum state.
    Phase noise is a typical assumption for optical fiber links \cite{PhysRevA.85.022340}.
\end{itemize}

\subsubsection{Coherent noise.}\label{sec:coherentnoise}

Coherent noise arises from systematic unitary perturbations, such as gate over-rotations or calibration inaccuracies leading to deviations of quantum states while maintaining its purity.
Qubit rotation operations are subject to inaccurate timing, if implemented based on time-dependent evolutions determining the rotation angle.
Consider the single-qubit rotation in \cref{eq:defRotation}, where the rotation angle $\phi$ determines the operation applied.
Assume the rotation speed of some implementation is given by $\omega$ and the target activation time is $t$ such that $\phi = \omega t$ and
\begin{equation}
    R_P(\phi) = R_P(\omega t) = e^{-i \omega t P/2}\;.
\end{equation}
If we have an imperfect implementation, for example due to some relative clock drift $\delta$, we effectively obtain a rotation of $\psi = \omega t (1+\delta)$ such that $\theta=\omega t \delta$ is the over-rotation error.
The corresponding imperfect implementation is
\begin{equation}
    R^\delta_P(\phi)=R_P(\psi) = e^{-i (\phi+\theta) P/2} = R_P(\phi)R_P(\theta)\;.
    \label{eq:over_rotation_paper}
\end{equation}

Such an implementation with a systematic offset at the clock (e.g., due to clock drift) results in systematic over-rotations across all rotational gates in a circuit.

Depending on the quantum hardware, the available gate set is limited and this kind of over-rotation might influence other gates as well.
Any circuit consisting of arbitrary logical gates will be transpiled into a hardware-aware equivalent circuit based on the limited gate set.
In Section~\ref{sec:transpilation}, we considered the intermediate transpilation step of decomposing any gate into a combination of qubit rotations.
In case of a hardware platform having standard rotational gates available, each gate observes systematic errors after being transpiled into equivalent rotational gates.
For controlled unitary gates, we obtain controlled over-rotations, e.g. for the CNOT/CX gate:
\begin{equation}
    \text{CNOT} = (\text{S} \otimes \text{I}) \text{CRX}(\pi) \;.
\end{equation}
For the Hadamard gate, there are several ways of implementation (up to a global phase):
\begin{align}\label{eq:defHadamardRotations}
    R_H(\pi)&=R_X(\pi)R_Y(\pi/2)=R_Z(\pi)R_Y(-\pi/2)\nonumber\\&=R_Y(\pi/2)R_Z(\pi)=R_Y(-\pi/2)R_X(\pi) \;.
\end{align}
If a Hadamard rotation itself is available as a gate, it can observe over-rotation around that axis.
Otherwise, implementing it through two other rotations, each of them might observe over-rotation, which might result in different noise effects due to the implementation order.

\begin{figure}[t]
\centering
\begin{subfigure}{0.23\textwidth}
    \includegraphics[trim={3.75cm 1.75cm 3.75cm 1.75cm},clip,width=\textwidth]{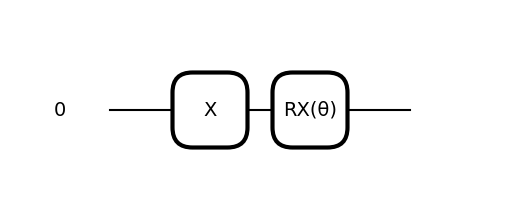}
    \caption{Pauli X gate.}
    \label{fig:overrotation-X}
\end{subfigure}
\hfill
\begin{subfigure}{0.23\textwidth}
    \includegraphics[trim={3.75cm 1.75cm 3.75cm 1.75cm},clip,width=\textwidth]{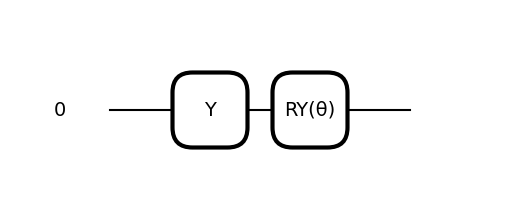}
    \caption{Pauli Y gate.}
    \label{fig:overrotation-Y}
\end{subfigure}
\hfill
\begin{subfigure}{0.23\textwidth}
    \includegraphics[trim={3.75cm 1.75cm 3.75cm 1.75cm},clip,width=\textwidth]{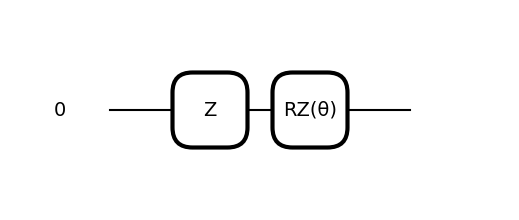}
    \caption{Pauli Z gate.}
    \label{fig:overrotation-Z}
\end{subfigure}
\hfill
\begin{subfigure}{0.23\textwidth}
    \includegraphics[trim={3.75cm 1.75cm 3.75cm 1.75cm},clip,width=\textwidth]{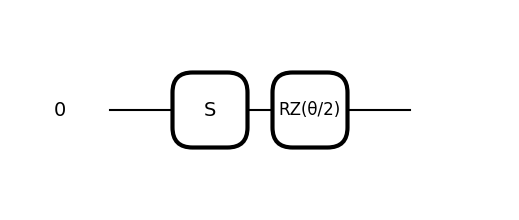}
    \caption{S gate.}
    \label{fig:overrotation-S}
\end{subfigure}

\vspace{1.5em}
\hfill
\begin{subfigure}{0.4\textwidth}
    \includegraphics[trim={3.75cm 0.5cm 3.75cm 1.75cm},clip,width=\textwidth]{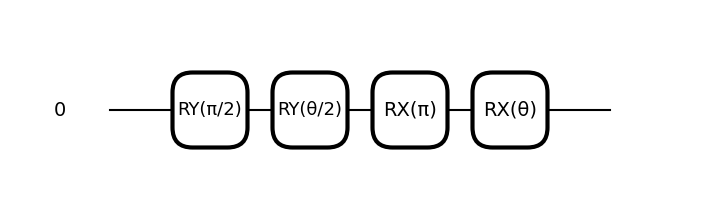}
    \caption{Hadamard gate.}
    \label{fig:overrotation-H}
\end{subfigure}
\hspace{1em}
\begin{subfigure}{0.4\textwidth}
    \includegraphics[trim={3.75cm 1.75cm 3.75cm 1.75cm},clip,width=\textwidth]{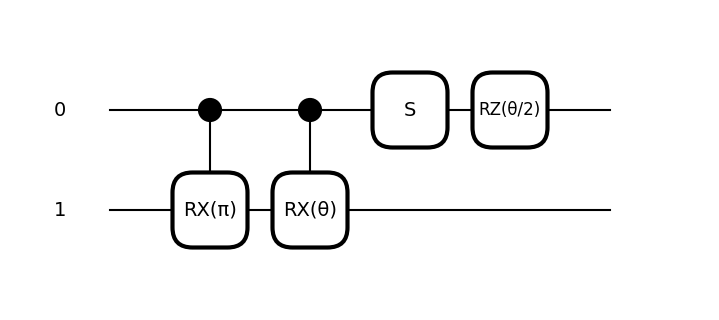}
    \caption{CNOT gate.}
    \label{fig:overrotation-CX}
\end{subfigure}
\hfill
\vspace{1.5em}

\begin{subfigure}{0.4\textwidth}
    \includegraphics[trim={3.75cm 1.75cm 3.75cm 1.75cm},clip,width=\textwidth]{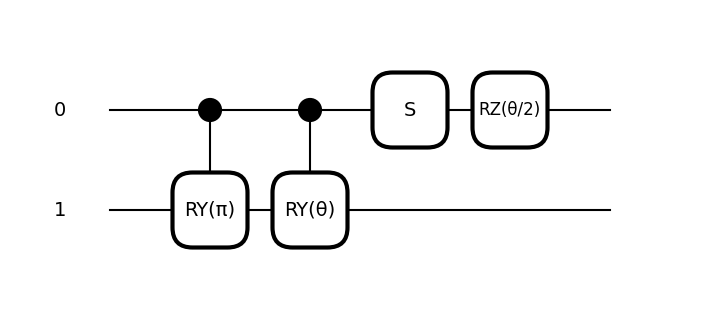}
    \caption{CY gate.}
    \label{fig:overrotation-CY}
\end{subfigure}
\hspace{1em}
\begin{subfigure}{0.4\textwidth}
    \includegraphics[trim={3.75cm 1.75cm 3.75cm 1.75cm},clip,width=\textwidth]{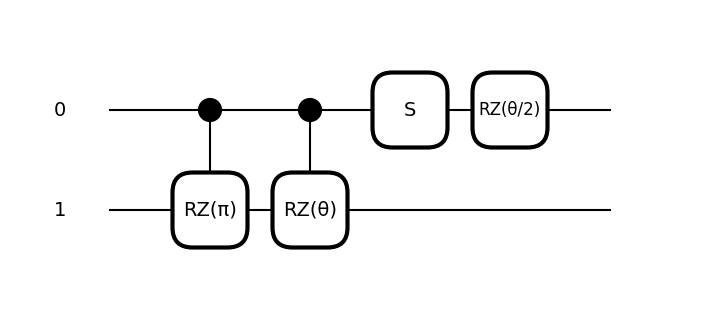}
    \caption{CZ gate.}
    \label{fig:overrotation-CZ}
\end{subfigure}
\caption{Implementation of a gate set using rotational gates with systematic over-rotations $\theta$ per rotation of $\pi$, constant across all gates.}
\label{fig:gates-overrotation}
\end{figure}
Coherent errors can accumulate with circuit depth and may interact with QEC differently than incoherent noise since it directly affects the error correction operations.
The coherent noise gates used in this work are provided in Figure~\ref{fig:gates-overrotation}.

\section{Simulation Setup}

In order to study the performance of error-correcting codes in second-generation quantum repeater links, we consider a simple two-segment repeater link and aim for the end-to-end link fidelity as the main indicator how well the entangled states are preserved.

\subsection{Two-segment Repeater Link}
\begin{figure}
    \centering
    \includegraphics[width=\linewidth]{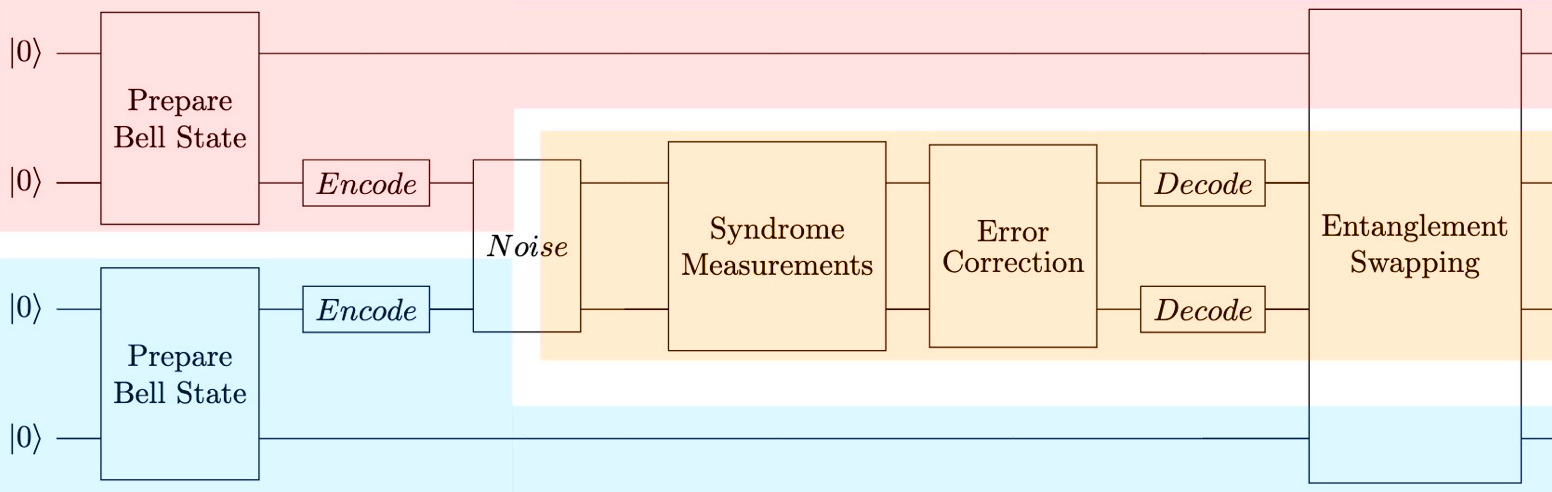}
    \caption{
    Schematic of a two-segment repeater link with QEC in place.
    During transmission of entangled qubits from Alice (red) and Bob (blue) to the intermediate repeater node (orange), a quantum error correcting code is employed.
    The block ``Noise'' accounts for any noise from memory, transmission and applied operations.
    }
    \label{fig:optimizing_fidelity_circuit}
\end{figure}

 We analyze the setup in Figure~\ref{fig:optimizing_fidelity_circuit} of a two-segment repeater with a single repeater node.
Alice and Bob each locally prepare a Bell pair and transmit one qubit to an intermediate repeater node, while the other qubit is stored locally in a quantum memory.
Before transmission of the traveling qubits, they are encoded into logical qubits using an error correcting code.
The transmission is modeled by applying single-qubit noisy channels to each physical qubit in the code block.
At the repeater, stabilizer syndromes are extracted for each received code block and decoded via a precomputed lookup table to apply correction operations.
The corrected logical qubits are then decoded back to single physical qubits.
Afterwards, an entanglement swapping operation is applied to these qubits to connect the two elementary links and extend entanglement between the end stations.
Mean fidelities are then calculated by comparison of the resulting shared (entangled) states with the initial Bell state.
The preparation of these initial Bell states is out of scope of this work.
Therefore, we assume perfect entangled states $\ket{\Phi^+}$ to be initially available to the nodes, independent of any error models considered afterwards.
Assuming the resulting state to be $\sigma$, we obtain the end-to-end fidelity as
\begin{equation}\label{eq:fidelity}
    F(\Phi^+,\sigma) = \bra{\Phi^+} \sigma \ket{\Phi^+} \;,
\end{equation}
which satisfies $0 \leq F(\Phi^+,\sigma) \leq 1$.

\subsection{Encoding and Decoding Circuits}

\begin{figure}[t]
\centering
\begin{subfigure}{1.0\textwidth}
    \includegraphics[width=1.0\linewidth]{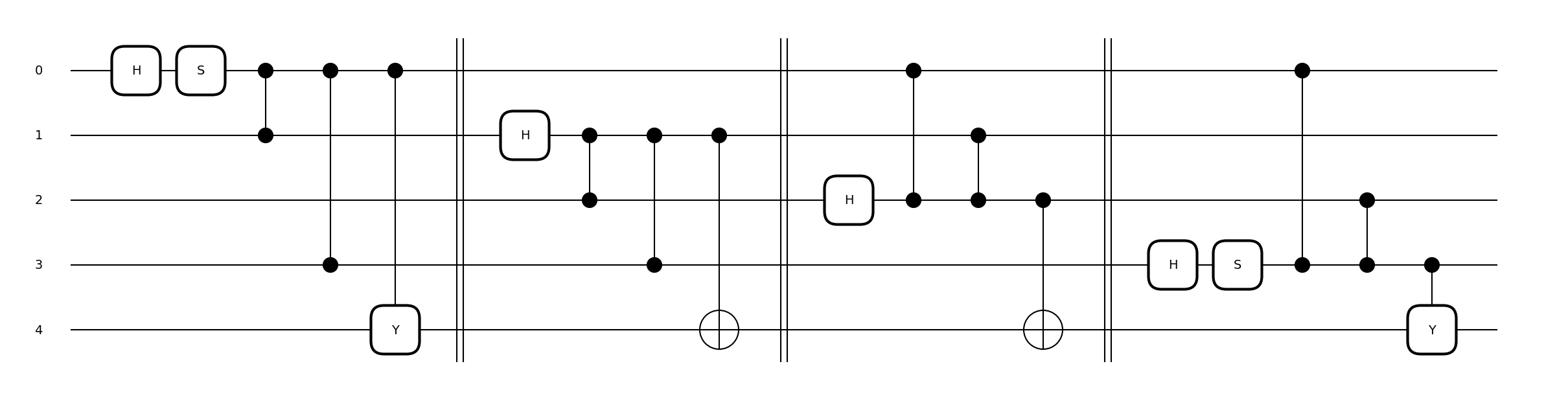}
    \caption{Encoding circuit for the $[[5,1,3]]$ code}
    \label{fig:5q_encoder}
\end{subfigure}
\begin{subfigure}{1.0\textwidth}
    \includegraphics[width=1.0\linewidth]{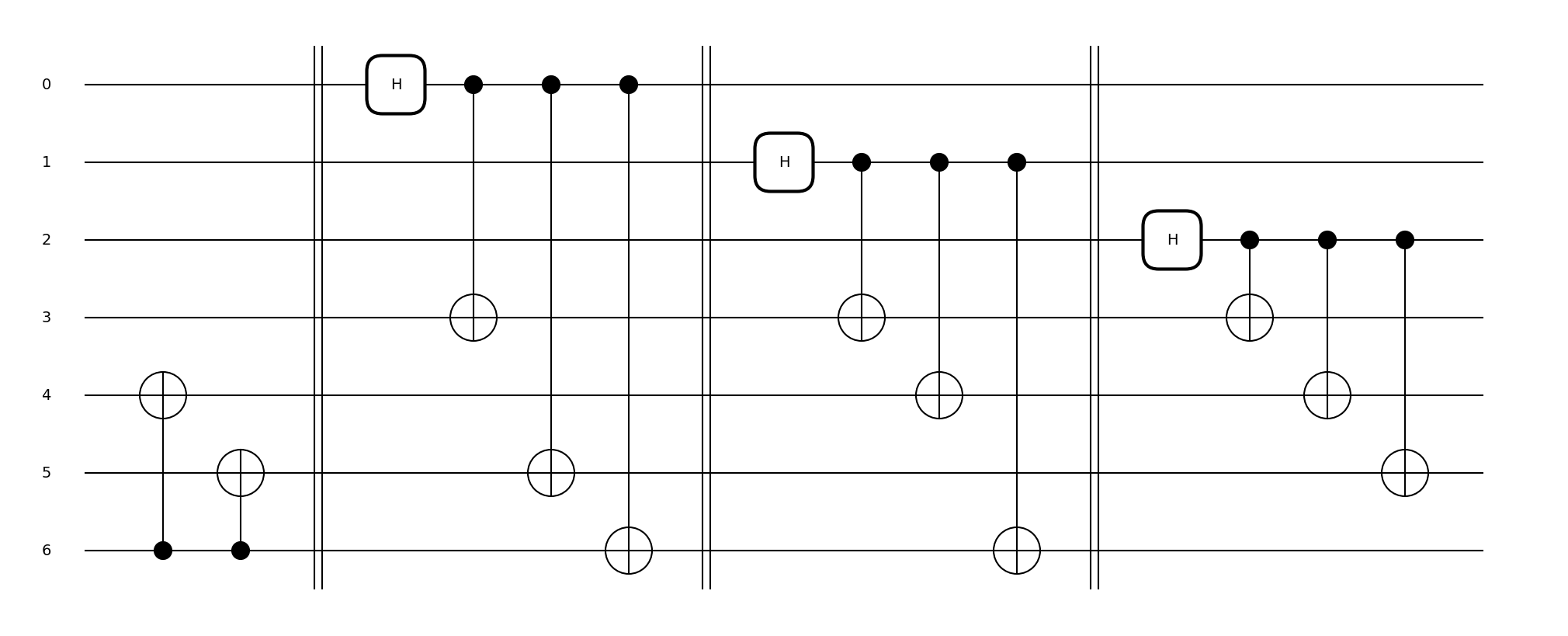}
    \caption{Encoding circuit for the $[[7,1,3]]$ code}
    \label{fig:7q_encoder}
\end{subfigure}
\caption{Example encoding circuits for the $[[5,1,3]]$ and $[[7,1,3]]$ code used to encode one half of a Bell pair (qubit 4 and 6, respectively). Any other qubit is initialized in state $\ket{0}$.}
\label{fig:encoders}
\end{figure}
In our analysis, we compare the performance of the $[[5,1,3]]$ and $[[7,1,3]]$ code to determine when which one is preferable.
The encoding circuits for these codes are generated using Gottesman-style encoder synthesis \cite{Gottesman2009AnIT}.
The resulting circuits used in our simulations are shown in Figure~\ref{fig:encoders}.
On the receiver side, syndrome measurements and correction operations are applied according to the code definitions in Section~\ref{sec:stabilizercodes}.
Afterwards, decoding is performed by applying the encoding circuits in reverse.

\subsection{Noise}
\label{sec:sim_noise}

Incoherent noise is introduced by the transmission channels from the distant nodes to the intermediate repeater node.
In order to analyze the code performance under a varying number of errors, we assume that only a subset of the physical qubits is affected by the noise while the rest observes noiseless channels.
The affected qubits are drawn uniformly at random and without replacement from the set of physical qubits.
This procedure is repeated in each simulation round (1000 per setting) and for each segment, while the number of noisy qubits is held fixed during the outer iteration of each simulation.
For the $[[5,1,3]]$ code, we end up with random subsets of up to five qubits that observe noise, while it is up to seven qubits for the $[[7,1,3]]$ code.

Coherent noise is introduced on the circuit-level implementations of the encoding, error-correction and decoding of the error-correcting codes.
For the implementation of the codes, we assume to have the noisy gate set in Figure~\ref{fig:gates-overrotation} available.
The error is modeled by a uniform over-rotation $\theta$ affecting all unitary gates in the circuit following the motivation and definition in Section~\ref{sec:coherentnoise}.
This model captures calibration errors in experimental platforms where gate implementations systematically deviate from their ideal parameters, for example due to an inaccurate clock.

\section{Simulation Results}
\label{sec:link_optimization}

In this section, we provide the analysis results comparing the performance of the $[[5,1,3]]$ and $[[7,1,3]]$ codes in a repeater link under different noise effects.
To quantify the degradation of the initial quantum states, we use the end-to-end fidelity of the repeater link defined in \cref{eq:fidelity} between the final shared state at the distant nodes and the target Bell state $\ket{\Phi^+}$.

\subsection{Incoherent Noise}

We prepare a logical Bell state locally utilizing the above codes and send one end of the entangled state to a distant station by transmitting the underlying physical qubits over transmission channels introducing incoherent noise.
The circuit implementations are assumed to be noiseless.
After receiving two entangled qubits at the intermediate repeater node, an entanglement swapping operation is applied.
The mean fidelity of the resulting shared Bell states is calculated.

\subsubsection{Depolarizing Noise.}

\begin{figure}
\centering
\begin{subfigure}{.82\textwidth}
    \includegraphics[width=1.0\linewidth]{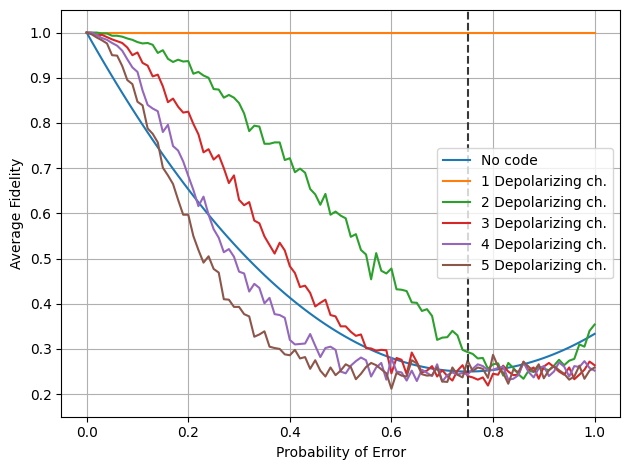}
    \caption{$[[5,1,3]]$ code}
    \label{fig:performance_depolarizing-5q}
\end{subfigure}
\begin{subfigure}{0.82\textwidth}
    \includegraphics[width=1.0\linewidth]{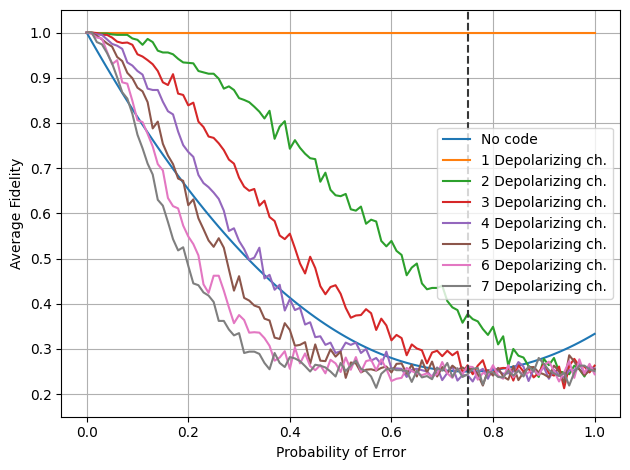}
    \caption{$[[7,1,3]]$ code}
    \label{fig:performance_depolarizing-7q}
\end{subfigure}
\caption{Performance trend of the codes under depolarizing noise with parameter $p$.
    The curve labeled ``No code'' corresponds to direct transmission over a noisy channel without quantum error correction.
    The dashed vertical line indicates maximal depolarizing noise.}
\label{fig:performance_depolarizing}
\end{figure}

\begin{figure}[t]
    \centering
    \includegraphics[width=1.0\linewidth]{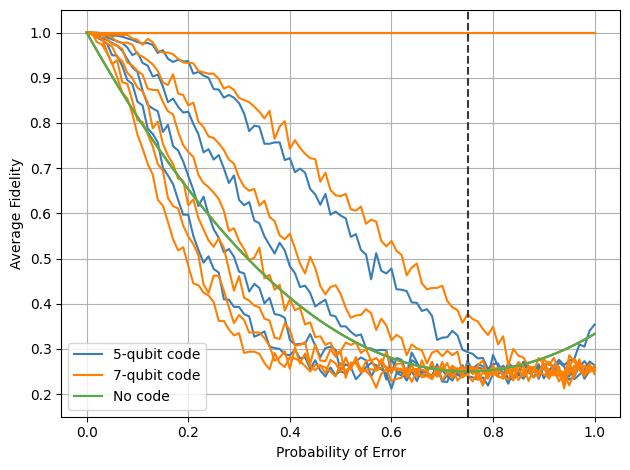}
    \caption{Combined performance trends of the codes under depolarizing noise with parameter $p$. Each curve corresponds to a different number of physical qubits affected by the noise.
    The curve labeled ``No code'' corresponds to direct transmission over a noisy channel without quantum error correction.
    The dashed vertical line indicates maximal depolarizing noise.
    }
    \label{fig:performance_depolarizing_combined}
\end{figure}

According to our noise model in Section~\ref{sec:sim_noise}, the number of depolarizing channels is controlled such that a subset of up to five (seven) of the physical qubits in the $[[5,1,3]]$ ($[[7,1,3]]$) code is affected by depolarizing noise.
The results are provided in Figure~\ref{fig:performance_depolarizing} and Figure~\ref{fig:performance_depolarizing_combined}.
On the vertical axis, we have the average fidelity, while the horizontal axes correspond to the depolarizing parameter $p$ of the transmission channels.
Each curve corresponds to a different number of noisy transmission channels affecting a subset of the physical qubits in the encoded block.
The large variations are due to limited computational power (1000 iterations).
The curves labeled ``No code'' indicate the computed fidelity obtained on a two-segment repeater link without an error correction code in place, thus having only one physical qubit traveling.
The dashed vertical line indicates maximal depolarizing noise.
In Figure~\ref{fig:performance_depolarizing_combined}, we joined the figures to make a better comparison of the two codes.
The blue curves are for the $[[5,1,3]]$ code while the orange ones are for the $[[7,1,3]]$ code.
In each case, the upper-most curve corresponds to a single noisy transmission and the lower-most to five and seven noisy transmissions, respectively.

As long as the noise is limited to single-qubit errors, both codes perfectly correct them in any case, as expected.
With more than one noisy channel, the fidelity reduces with increasing error probability and as the number of affected qubits increases, the fidelity decreases faster.
Particularly in the cases where all qubits are affected by the noise, both codes show an advantage compared to employing no code only if the error probability is below some threshold at around $0.15$ for the $[[5,1,3]]$ code and $0.1$ for the $[[7,1,3]]$ code.
Any increase in fidelity after $\frac{3}{4}$ can be explained by the fact that this value introduces maximal depolarization, as pointed out in Section~\ref{sec:incoherentnoise}.
Therefore, the qubit states observing $p>\frac{3}{4}$ are not maximally mixed anymore.
In the direct comparison of the two codes under two/three noisy channels, the $[[7,1,3]]$ code shows a small advantage compared to the $[[5,1,3]]$ code.
\begin{figure}
\centering
\begin{subfigure}{.82\textwidth}
    \includegraphics[width=1.0\linewidth]{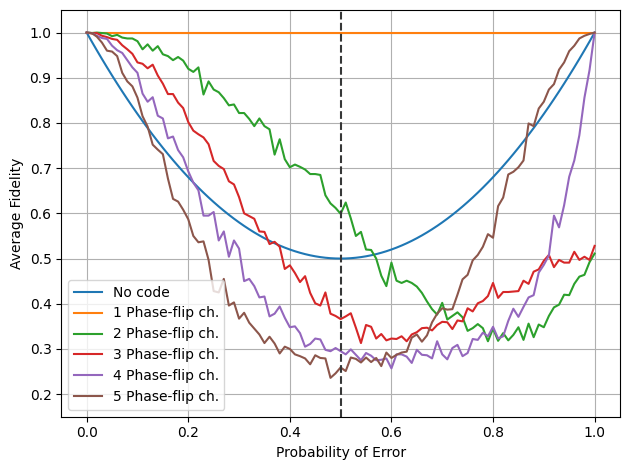}
    \caption{$[[5,1,3]]$ code}
    \label{fig:performance_dephasing-5q}
\end{subfigure}
\begin{subfigure}{.82\textwidth}
    \includegraphics[width=1.0\linewidth]{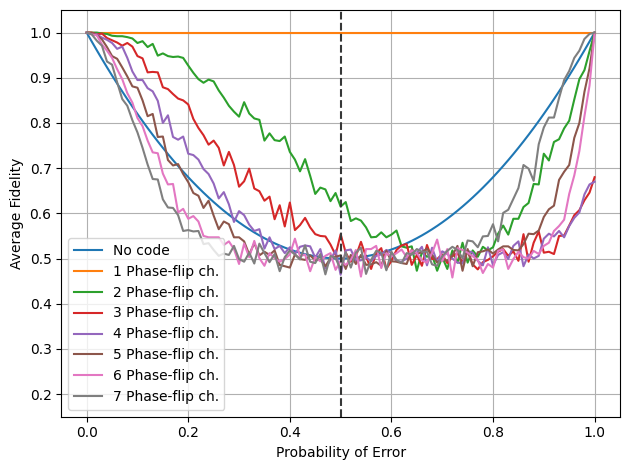}
    \caption{$[[7,1,3]]$ code}
    \label{fig:performance_dephasing-7q}
\end{subfigure}
\caption{Performance trend of the codes under phase-flip noise with parameter $p$.
    The curve labeled ``No code'' corresponds to direct transmission over a noisy channel without quantum error correction.
    The dashed vertical line indicates maximal dephasing noise.}
\label{fig:performance_dephasing}
\end{figure}
With an increasing number of affected qubits, this advantage vanishes and eventually, at the maximum of seven qubits, the fidelity drops below that one of the $[[5,1,3]]$ code with five noisy qubits.
At very small error probabilities, our data does not allow a proper comparison of the codes.

\subsubsection{Phase-flip Noise.}

\begin{figure}[t]
    \centering
    \includegraphics[width=1.0\linewidth]{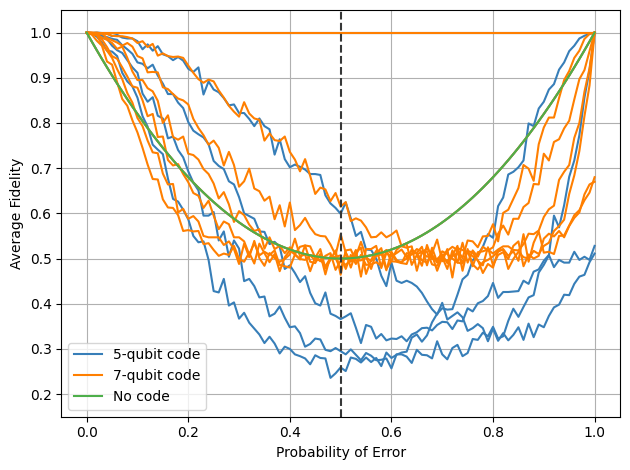}
    \caption{Combined performance trends of the codes under phase-flip noise with parameter $p$. Each curve corresponds to a different number of physical qubits affected by the noise.
    The curve labeled ``No code'' corresponds to direct transmission over a noisy channel without quantum error correction.
    The dashed vertical line indicates maximal dephasing noise.
    }
    \label{fig:performance_dephasing_combined}
\end{figure}

Next, we assume that the transmission channel introduces phase-flip noise to a subset of the physical qubits of the code, according to our noise model in Section~\ref{sec:sim_noise}.
The results are provided in Figure~\ref{fig:performance_dephasing} and Figure~\ref{fig:performance_dephasing_combined} in the same way as previously for the depolarizing noise.
For phase-flip noise, we effectively need to only consider error probabilities below $0.5$ (dashed vertical line), which corresponds to maximum dephasing as discussed in Section~\ref{sec:incoherentnoise}.
Again, both codes correct single qubit errors as expected.
For more than one noisy qubit, we observe that the fidelity drops to at most $0.5$ with the $[[7,1,3]]$ code while the $[[5,1,3]]$ code yields significantly lower fidelities, such that a clear separation between the codes emerges at higher error probabilities.

Comparing the codes at low phase-flip noise on subsets of their qubits, both codes perform comparable.
Similar to the depolarizing noise, employing no code outperforms both codes above some threshold.
However, with all physical qubits being noisy, the fidelity with the $[[7,1,3]]$ code initially drops faster below the no-code performance, but ends up at fidelity $0.5$ while the $[[5,1,3]]$ code first keeps up with employing no code before approaching comparably low fidelity.

\subsection{Coherent Noise}

\begin{figure}[t]
    \centering
    \includegraphics[width=1.0\linewidth]{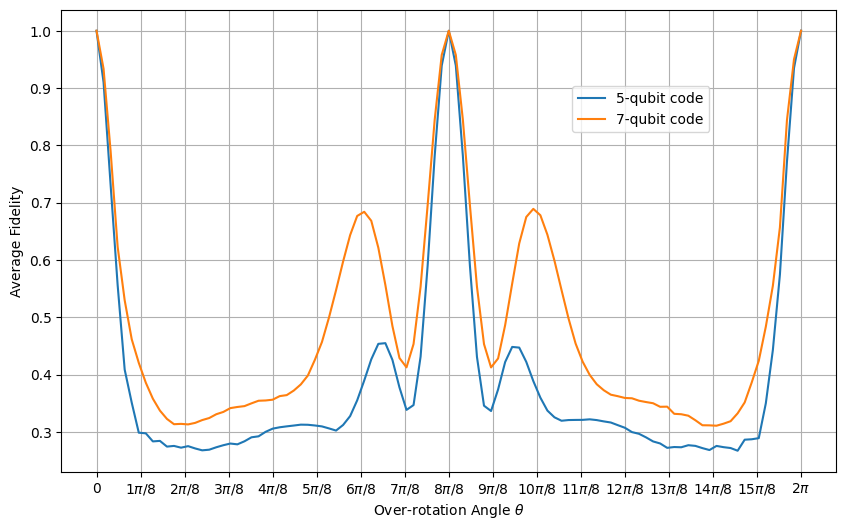}
    \caption{Comparison of the $[[5,1,3]]$ and $[[7,1,3]]$ codes under coherent over-rotation noise on all gates, e.g., due to clock drift.}
    \label{fig:performance_coherent}
\end{figure}
For the coherent noise analysis, we start with noiseless transmission of qubits between nodes, but the circuit implementations of the employed codes observe coherent noise in the form of gate over-rotations, as defined in Section~\ref{sec:sim_noise}.
Entanglement generation at the transmitters and entanglement swapping at the intermediate node are assumed noiseless.
The mean fidelity of the resulting shared Bell states is calculated.
The code comparison yields Figure~\ref{fig:performance_coherent}.
\begin{figure}
\centering
\begin{subfigure}{1.0\textwidth}
    \includegraphics[width=\textwidth]{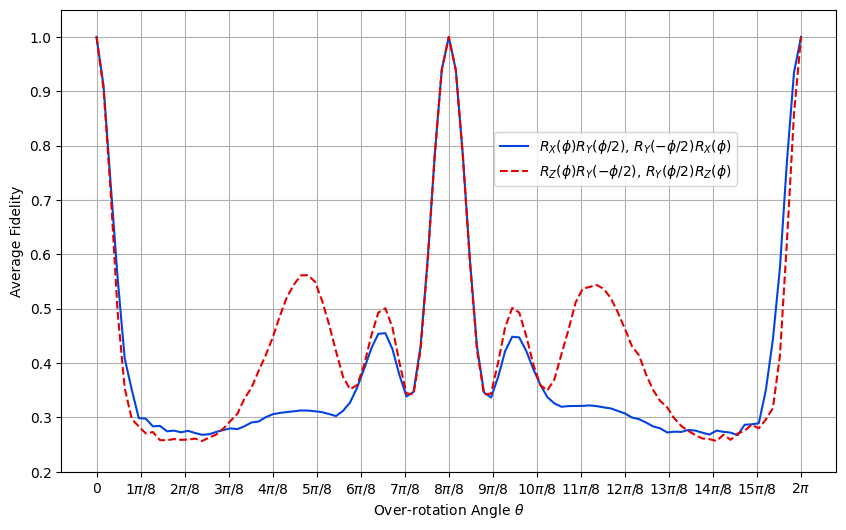}
    \caption{$[[5,1,3]]$ code}
    \label{fig:overrotation-H5}
\end{subfigure}
\begin{subfigure}{1.0\textwidth}
    \includegraphics[width=\textwidth]{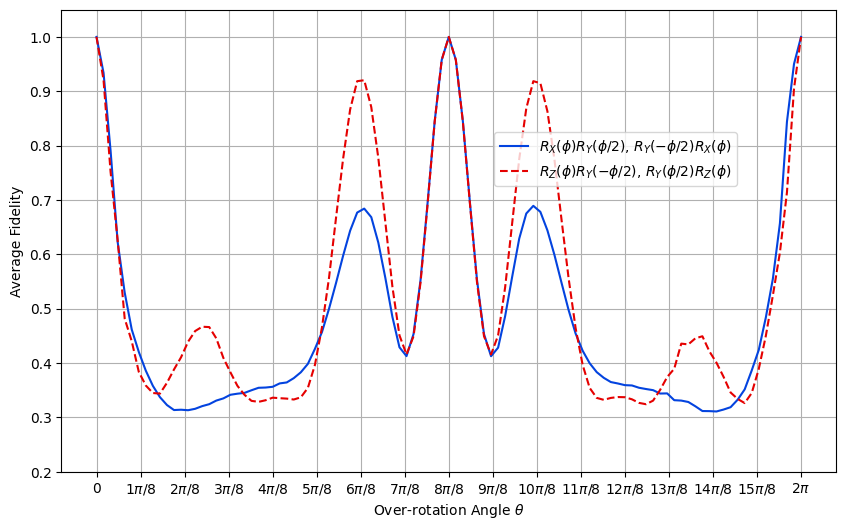}
    \caption{$[[7,1,3]]$ code}
    \label{fig:overrotation-H7}
\end{subfigure}        
\caption{Comparison of effects due to different Hadamard gate implementations according to \cref{eq:defHadamardRotations}.}
\label{fig:gates-overrotation-H-comparison}
\end{figure}

The first observation in both codes is its symmetric behavior regarding the end-to-end fidelity.
In case of a systematic error affecting all gates, only the absolute value of over-rotation is relevant.
Further, an over-rotation of $\pi$ yields perfect fidelity again.
This indicates that the error-correction in both codes is similarly effective with inverted gates.
Comparing the performance of the two codes, we observe that the $[[7,1,3]]$ code is constantly outperforming the $[[5,1,3]]$ code, while this effect is not too strong for small over-rotations.
In that regime, the states quickly lose fidelity with both codes as the over-rotation increases.

With the definition of the Hadamard gate in \cref{eq:defHadamardRotations}, we discussed that the choice of gate implementation can affect the observed error.
Figure~\ref{fig:gates-overrotation-H-comparison} provides a comparison of the effects from different Hadamard gate implementations within the considered repeater link.
We observe that in both codes, the implementations from \cref{eq:defHadamardRotations} result in pairwise identical behaviors.
The blue curves correspond to the results in Figure~\ref{fig:gates-overrotation}.
Regarding the considered coherent over-rotation, the implementation choice is a trade-off in the performance, since there is no superior implementation but depending on the observed over-rotation, one or the other implementation provides significantly higher fidelities.
Small over-rotation errors in the implementations result in similar effects on the fidelity, with a slight advantage in stability for the implementations $H=R_X(\phi)R_Y(\phi/2)=R_Y(-\phi/2)R_X(\phi)$.

\subsection{Combined Noise}

\begin{figure}
\centering
\begin{subfigure}{.82\textwidth}
    \includegraphics[width=1.0\linewidth]{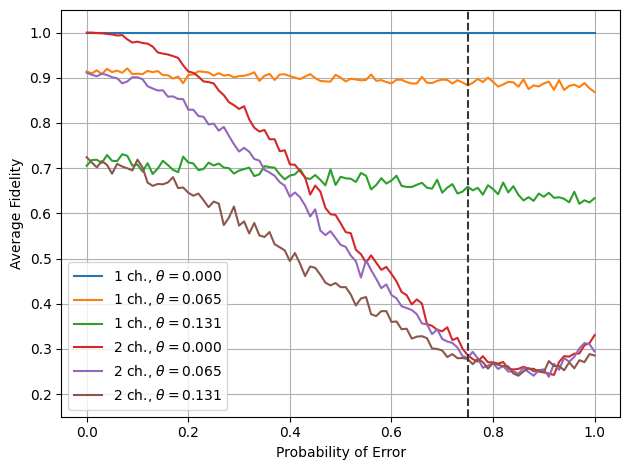}
    \caption{$[[5,1,3]]$ code}
    \label{fig:performance_combined-5q}
\end{subfigure}
\begin{subfigure}{.82\textwidth}
    \includegraphics[width=1.0\linewidth]{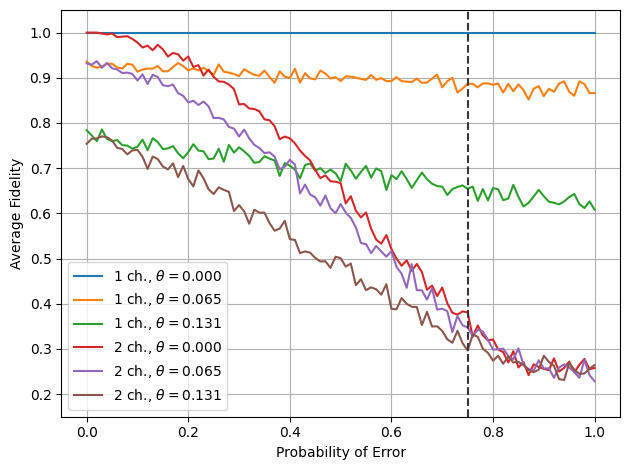}
    \caption{$[[7,1,3]]$ code}
    \label{fig:performance_combined-7q}
\end{subfigure}
\caption{Comparison of the $[[5,1,3]]$ and $[[7,1,3]]$ codes under one or two depolarizing channels and gate over-rotation of $\theta=\{0, \frac{\pi}{48}, \frac{\pi}{24}\}$.}
\label{fig:performance_combined}
\end{figure}

Finally, we consider a simple mixed noise model that combines incoherent noise with coherent errors.
Particularly, we consider the case of coherent gate over-rotations as in the previous sections and add either one or two depolarizing channels acting under the transmission of physical qubits.
We limit the analysis to these two cases since the previous sections have shown that a single qubit depolarizing error is expected to be perfectly corrected while more than one always shows the similar characteristic of reduced fidelity with increasing noise.
Now, combining these cases with over-rotation errors, we ask the question whether any small calibration errors impact the error correction performance with respect to depolarizing errors or whether these two contributions can be considered independent.
Therefore, the values for the over-rotation were chosen small in the sense that the fidelity under pure coherent errors does not drop below $0.7$. Varying the depolarizing noise contribution, we obtain Figure~\ref{fig:performance_combined}.

First, compare the curves without over-rotation $(\theta=0)$ to these including an over-rotation.
We observe a performance degradation due to this additional error.
However, the general behavior with respect to the depolarizing error does not change significantly for any number of affected channels.
In the case of single-qubit depolarizing errors, the fidelity remains on a similar level as observed in Figure~\ref{fig:performance_coherent} with over-rotation only.
We observe a moderate decrease with increasing depolarizing error probability, indicating that the code's error correction capabilities degrade slightly.
With two depolarizing channels, the fidelity drops nonetheless such that an additional degradation is not noticeable in our data.

\begin{figure}[t]
    \centering
    \includegraphics[width=1.0\linewidth]{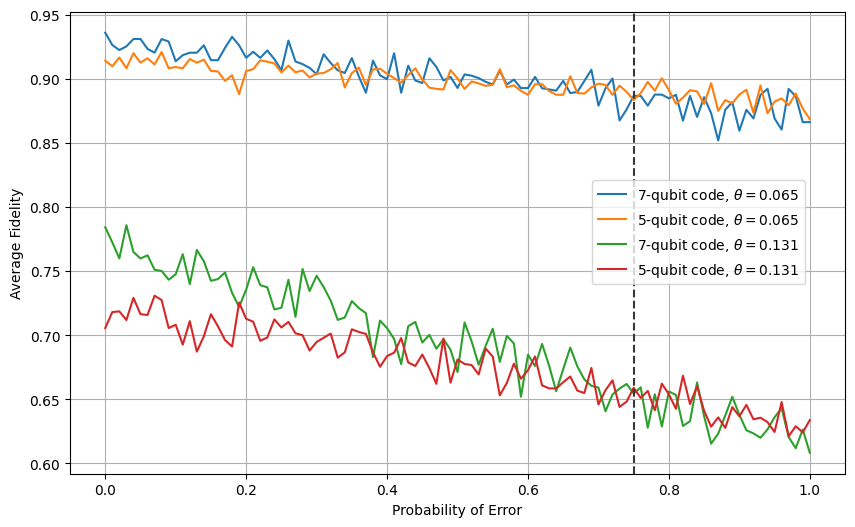}
    \caption{Comparison of the $[[5,1,3]]$ and $[[7,1,3]]$ codes under one depolarizing channel and small gate over-rotations $\theta$.}
    \label{fig:performance_combined_detail}
\end{figure}

In Figure~\ref{fig:performance_combined_detail}, we take a closer look at the code performance at a single depolarizing channel.
Our data indicates that the fidelity constantly decreases in both codes as the depolarizing probability increases.
This suggests that the over-rotation affects the correction capabilities of the code moderately, but increasingly with larger over-rotation angles.
Further, it seems that both codes achieve the same fidelity as the depolarizing probability approaches one.
As the $[[7,1,3]]$ code starts higher under pure over-rotations, the performance degradation is larger for that code.

\section{Discussion}

\subsection{Noise Effects}

\subsubsection{Incoherent Noise.}

Depending on the number of physical qubits affected by the noise, we observe a significantly higher fidelity compared to using no code, even at maximum depolarization or dephasing.
This is possible due to the employed noise model.
In practice, it is reasonable to assume that all qubits observe the same channel, while we assumed some qubits to remain noiseless in order to analyze the error-correction capabilities separately.
Due to these noiseless qubits, the fidelity might remain higher.
The low fidelity obtained with the $[[5,1,3]]$ code under phase-flip noise shows that this is not necessarily the case.
The fidelity below $0.5$ implies that the resulting states are not only a mixture of two Bell states with a difference in the phase.
Rather, further unwanted states occur after applying the error correction.

\subsubsection{Coherent Noise.}

We considered systematic gate over-rotations and observed a reduced fidelity of the entangled state.
A question regarding our results is whether this fidelity is due to noise introduced by erroneous operations applied by the error correction circuit or is it solely the result of the over-rotation, such that the obtained state remains a pure state?
In the latter case, adding a systematic correction operation could account for the coherent noise effects.

In general, coherent errors on the gates can accumulate with circuit depth, which in the present case depends on the encoder/decoder and syndrome-extraction circuits.
We consider the total gate count.
For the $[[5,1,3]]$ code, it is 18 gates (16 Hadamard or controlled) in the encoding/decoding circuit and 24 gates (24 Hadamard or controlled) in the syndrome measurement circuit.
Ignoring the correction operation that is potentially applied, we count a total of 116 physical gates, where each Hadamard and controlled gate is implemented by two rotational gates.
For the $[[7,1,3]]$ code, it is 14 gates (14 Hadamard or controlled) in the encoding/decoding and 36 gates (36 Hadamard or controlled) in the syndrome measurement, totaling to 128 physical gates.
This shows that the gate count of the two considered codes is of the same magnitude, with a slight advantage for the $[[5,1,3]]$ code.
Therefore, this analysis is not sufficient to explain the slightly worse performance of the $[[5,1,3]]$ code at small over-rotations.
Larger over-rotations might influence the implemented logic of the circuits, resulting in significant differences independent of circuit depth, but certainly depending on the exact circuit implementation.

Studying the effect of the Hadamard gate implementation, we conclude that the physical gate implementation has significant effects on how an over-rotation error affects the outcome.
However, assuming over-rotations due to calibration errors like a clock drift, we expect small over-rotations, where the implementation choice does not have significant effects on the overall code performance.
Regarding any other gate, it is always possible to implement them by a positive or negative rotation, which would result in corresponding over-rotation errors.
By a wise implementation choice for each gate, this could be used to cancel out errors.
For example, the gates in our encoding and decoding circuits could be implemented to rotate in the opposite direction.
To evaluate the effects of these measures, a detailed study of the code circuits is necessary.

\subsubsection{Combined Noise.}

Combining over-rotations with depolarizing noise, we found that their effects are largely independent, although they do influence each other to some extent.
Additional over-rotation degrades the performance of both codes, while the correction of single-qubit errors in general remains still possible, with an added noise due to the over-rotation.
However, both codes observe a small decrease in fidelity as the depolarizing noise increases, showing the existence of moderate cross-effects.

\subsection{Code Selection}

Regarding the code selection in a second-generation quantum repeater, we can summarize our findings as follows:
The code choice depends on the dominant physical noise.
If over-rotation noise is dominant, the $[[7,1,3]]$ code is the slightly better option.
In the low-noise regime of depolarizing and phase-flip noise, the $[[5,1,3]]$ code is preferable due to potentially better performance and lower resource requirements (number of qubits).
With increasing noise, however, this code renders useless under phase-flip noise, while the fidelity at the $[[7,1,3]]$ remains higher but drops as well.
In higher noise regimes, it turns out that the best strategy is using no code, having less qubits susceptible to noise and no additional over-rotation error due to encoding and decoding operations.
As the number of physical qubits increases, the probability of observing uncorrectable multi-qubit errors increases, and additional gates accumulate even small over-rotation errors during processing.
This results in a specific threshold regarding the observed noise, where the logical error exceeds the physical one, hence supporting the idea of using the smaller $[[5, 1, 3]]$ code at low noise levels and dropping error correction entirely as noise increases.
If it was possible to guarantee only a subset of the code qubits being affected by the noise, the codes perform significantly better.
Yet, such a case would raise the question of why such a noiseless channel cannot be used for transmission without employing a code.

\section{Conclusions}
\label{sec:conclusions}

This work evaluated the performance of different error correcting codes in a second-generation quantum repeater link under various noise conditions.
To that extent, we simulated a two-segment repeater link with the $[[5,1,3]]$ and $[[7,1,3]]$ stabilizer codes in a hardware-aware gate set.
We incorporated noise on the transmission channels of the physical qubits and at the circuit level, classified into incoherent and coherent components.
Considering the end-to-end entangled state fidelity, we were able to quantify the effects of different noise on the code performance and identify the best code choice for specific noise regimes.
While high noise regimes suggest employing no error correction codes, an advantage is possible for low noise levels.
In that case, the smaller $[[5,1,3]]$ code slightly outperforms the $[[7,1,3]]$ code at depolarizing and phase-flip noise.
As soon as the coherent over-rotations on the gates, for example due to a clock drift, dominate the error, the $[[7,1,3]]$ code performs better.
In the case of simultaneous small over-rotations and depolarizing noise, our simulations indicate only moderate cross-effects on the code performance.
Further, it is shown that the choice for the physical implementation of gates has a significant influence on the code performance as well.

These findings are limited to the two codes considered and their circuit implementations, which were not further optimized.
Without analytical backing, the findings cannot be directly generalized to other implementations, larger codes or other code families.

Possible future work includes incorporating heterogeneous and time-varying noise models, and exploring adaptive coding strategies that select codes and decoding rules based on real-time device calibration data.
To understand the obtained simulation results in more detail, investigating the circuit-level implementations of the codes is required.

\sloppy
\section*{Acknowledgments}

The authors gratefully acknowledge financial support from the \textbf{QF2 project} (En\-tang\-lement-assisted communication), funded by the Deutsche Forschungsgemein\-schaft (DFG, German Research Foundation) under Germany’s Excellence Strategy – EXC 2123/2 QuantumFrontiers – 390837967; the \textbf{QSTARS project} (Post-quantum cryptography, entanglement-assisted communication, identification, oblivious transfer), funded by the Bundesministerium für Forschung, Technologie und Raumfahrt (BMFTR) under grant 16KIS2602; the \textbf{QR.N project} (Quantum repeaters, hybrid quantum networks, super-dense coding), funded by the BMFTR under grant 16KIS2196; the \textbf{6G-life project} (Identification: information-theoretic bounds and identification codes), funded by the BMFTR under grant 16KIS2415; the \textbf{QUIET project} (Identification codes using quantum resources), funded by the BMFTR under grant 16KISQ0170; the \textbf{QD-CamNetz project} (Hybrid quantum and classical networks, oblivious transfer), funded by the BMFTR under grant 16KISQ169; the \textbf{Q-TREX project} (Malicious users in networks), funded by the BMFTR under grant 16KISR038; and the \textbf{FOCAL project} (CV-QKD via satellites), funded by the European Union under the Marie Skłodowska-Curie Actions (MSCA) within Horizon Europe.

\bibliographystyle{spbasic}
\bibliography{references}

@article{RevModPhys.95.045006,
    title = {Quantum repeaters: From quantum networks to the quantum internet},
    author = {Azuma, Koji and Economou, Sophia E. and Elkouss, David and Hilaire, Paul and Jiang, Liang and Lo, Hoi-Kwong and Tzitrin, Ilan},
    journal = {Rev. Mod. Phys.},
    volume = {95},
    issue = {4},
    pages = {045006},
    numpages = {66},
    year = {2023},
    month = {Dec},
    publisher = {American Physical Society},
    doi = {10.1103/RevModPhys.95.045006}
}

@article{PhysRevLett.81.5932,
    title = {Quantum Repeaters: The Role of Imperfect Local Operations in Quantum Communication},
    author = {Briegel, H.-J. and D\"ur, W. and Cirac, J. I. and Zoller, P.},
    journal = {Phys. Rev. Lett.},
    volume = {81},
    issue = {26},
    pages = {5932--5935},
    numpages = {0},
    year = {1998},
    month = {Dec},
    publisher = {American Physical Society},
    doi = {10.1103/PhysRevLett.81.5932}
}

@article{PhysRevA.54.1098,
    title = {Good quantum error-correcting codes exist},
    author = {Calderbank, A. R. and Shor, Peter W.},
    journal = {Phys. Rev. A},
    volume = {54},
    issue = {2},
    pages = {1098--1105},
    numpages = {0},
    year = {1996},
    month = {Aug},
    publisher = {American Physical Society},
    doi = {10.1103/PhysRevA.54.1098}
}

@article{Fan2025CalibratingQG,
    title={Calibrating quantum gates up to 52 qubits in a superconducting processor},
    author={Daojin Fan and Guoding Liu and Shaowei Li and Ming Gong and Dachao Wu and Yiming Zhang and Chen Zha and Fusheng Chen and Sirui Cao and Yangsen Ye and Qingling Zhu and Chong Ying and Shaojun Guo and Haoran Qian and Yulin Wu and Hui Deng and Gang Wu and Cheng-Zhi Peng and Xiongfeng Ma and Xiaobo Zhu and Jian-Wei Pan},
    journal={npj Quantum Inf},
    volume={11},
    pages={33},
    year={2025},
    doi={10.1038/s41534-025-00983-5}
}

@article{PhysRevA.80.042308,
    title = {Entanglement purification with double selection},
    author = {Fujii, Keisuke and Yamamoto, Katsuji},
    journal = {Phys. Rev. A},
    volume = {80},
    issue = {4},
    pages = {042308},
    numpages = {9},
    year = {2009},
    month = {Oct},
    publisher = {American Physical Society},
    doi = {10.1103/PhysRevA.80.042308}
}

@book{gottesman1997stabilizer,
    title={Stabilizer codes and quantum error correction},
    author={Gottesman, Daniel},
    year={1997},
    publisher={California Institute of Technology}
}

@inproceedings{Gottesman2009AnIT,
    title={An introduction to quantum error correction and fault-tolerant quantum computation},
    author={Gottesman, Daniel},
    booktitle={Quantum information science and its contributions to mathematics, Proceedings of Symposia in Applied Mathematics},
    volume={68},
    pages={13--58},
    year={2010},
    doi={10.48550/arXiv.0904.2557}
}

@article{Jiang_Taylor_Nemoto_Munro_VanMeter_Lukin_2009,
    title={Quantum repeater with encoding},
    volume={79},
    DOI={10.1103/PhysRevA.79.032325},
    number={3},
    journal={Physical Review A},
    publisher={American Physical Society},
    author={Jiang, Liang and Taylor, J. M. and Nemoto, Kae and Munro, W. J. and Van Meter, Rodney and Lukin, M. D.}, year={2009},
    month=mar,
    pages={032325}
}

@article{Laflamme1996PerfectQE,
    title = {Perfect Quantum Error Correcting Code},
    author = {Laflamme, Raymond and Miquel, Cesar and Paz, Juan Pablo and Zurek, Wojciech Hubert},
    journal = {Phys. Rev. Lett.},
    volume = {77},
    issue = {1},
    pages = {198--201},
    numpages = {0},
    year = {1996},
    month = {Jul},
    publisher = {American Physical Society},
    doi = {10.1103/PhysRevLett.77.198}
}

@article{Loock2019ExtendingQL,
    author = {van Loock, Peter and Alt, Wolfgang and Becher, Christoph and Benson, Oliver and Boche, Holger and Deppe, Christian and Eschner, Jürgen and Höfling, Sven and Meschede, Dieter and Michler, Peter and Schmidt, Frank and Weinfurter, Harald},
    title = {Extending Quantum Links: Modules for Fiber- and Memory-Based Quantum Repeaters},
    journal = {Advanced Quantum Technologies},
    volume = {3},
    number = {11},
    pages = {1900141},
    doi = {10.1002/qute.201900141},
    year = {2020}
}

@article{Muralidharan2015OptimalAF,
    author={Muralidharan, Sreraman
    and Li, Linshu
    and Kim, Jungsang
    and L{\"u}tkenhaus, Norbert
    and Lukin, Mikhail D.
    and Jiang, Liang},
    title={Optimal architectures for long distance quantum communication},
    journal={Scientific Reports},
    year={2016},
    month={Feb},
    day={15},
    volume={6},
    number={1},
    pages={20463},
    issn={2045-2322},
    doi={10.1038/srep20463}
}

@book{Chuang2020QuantumCA,
    place={Cambridge},
    title={Quantum Computation and Quantum Information: 10th Anniversary Edition},
    publisher={Cambridge University Press},
    author={Nielsen, Michael A. and Chuang, Isaac L.},
    year={2010}
}

@article{PhysRevA.85.022340,
    title = {Dynamical decoupling in optical fibers: Preserving polarization qubits from birefringent dephasing},
    author = {Roy Bardhan, Bhaskar and Anisimov, Petr M. and Gupta, Manish K. and Brown, Katherine L. and Jones, N. Cody and Lee, Hwang and Dowling, Jonathan P.},
    journal = {Phys. Rev. A},
    volume = {85},
    issue = {2},
    pages = {022340},
    numpages = {7},
    year = {2012},
    month = {Feb},
    publisher = {American Physical Society},
    doi = {10.1103/PhysRevA.85.022340}
}

@article{RevModPhys.83.33,
    title = {Quantum repeaters based on atomic ensembles and linear optics},
    author = {Sangouard, Nicolas and Simon, Christoph and de Riedmatten, Hugues and Gisin, Nicolas},
    journal = {Rev. Mod. Phys.},
    volume = {83},
    issue = {1},
    pages = {33--80},
    numpages = {0},
    year = {2011},
    month = {Mar},
    publisher = {American Physical Society},
    doi = {10.1103/RevModPhys.83.33}
}

@article{PhysRevLett.77.793,
    title = {Error Correcting Codes in Quantum Theory},
    author = {Steane, A. M.},
    journal = {Phys. Rev. Lett.},
    volume = {77},
    issue = {5},
    pages = {793--797},
    numpages = {0},
    year = {1996},
    month = {Jul},
    publisher = {American Physical Society},
    doi = {10.1103/PhysRevLett.77.793}
}

@article{Zwerger_Briegel_Dür_2014,
    title={Hybrid architecture for encoded measurement-based quantum computation},
    volume={4},
    ISSN={2045-2322},
    DOI={10.1038/srep05364},
    number={1},
    journal={Scientific Reports},
    author={Zwerger, M. and Briegel, H. J. and Dür, W.},
    year={2014},
    month=jun,
    pages={5364}
}
\end{document}